\documentstyle[12pt,amsfonts]{article}
\begin{document}
\hfill solv-int/9705006\footnote{NTZ-Preprint 24/95, Leipzig, 1995, 
to appear in J. of Applied Mathematics (ZAMP)}

\begin{center}{\large Investigation of dynamical systems using tools of the 
theory of invariants and projective geometry}\\[12pt]

{L A Bordag\dag ~~ and V S Dryuma\ddag}

{\dag\ Mathematisches Institut, Universit\"at
Leipzig, Augustusplatz 10, 04109 Leipzig, Germany}\\
{\ddag\ Institute of Mathematics, Academy of Science, RM, Kishinev, 277028, 
Moldova}
\end{center}

\begin{abstract}
The investigation of nonlinear dynamical systems of the type
$$\dot{x}=P(x,y,z),\dot{y}=Q(x,y,z),\dot{z}=R(x,y,z)$$
by means of reduction to some ordinary differential equations of
the second order in the form
\[y''+a_1(x,y)y'^3+3a_2(x,y)y'^2+3a_3(x,y)y'+a_4(x,y)=0\]
is done. The main backbone of this investigation was provided by
the theory of invariants developed by S. Lie, R. Liouville and A.
Tresse at the end of the 19$^{\rm th}$ century and the
projective geometry of E. Cartan. In our work two, in some sense
supplementary, systems are considered. The first one is the
Lorenz system
$$\dot{x}=\sigma (y-x),\dot{y}=rx-y-zx\,, \dot{z}=xy-bz\,$$
where $\sigma ,r,b$ are parameters and the second one is the
R\"o\ss ler system
$$\dot{x}=-y-z,\dot{y}=x+ay\,, \dot{z}=b+xz-cz\,$$
where $a,b$ and $c$ are parameters.  The invarinats for the
ordinary differential equations, which correspond to the systems
mentioned abouve, are evaluated. The connection of values of the invariants 
with characteristics of dynamical systems is established. 
\end{abstract}
\break
\section{Introduction}

The first broad investigation of invariants of the second order
differential equation
\begin{equation}y''=f(x,y,y')\label{1}\end{equation}
under some general point transformations
\begin{eqnarray}
x&=&\xi (u,v)\,, \nonumber\\
y&=&\eta (u,v)\,, \label{2}\end{eqnarray} 
where
$\xi$ and $\eta$ are arbitrary smooth functions was done in the
work of R. Liouville \cite{Liouville}.  He found some series of
absolute and semi-invariants and discovered one procedure to
build other invariants of higher weights.  A further
consideration of the same problem from the point of view of
infinitesimal transformations was introduced by S. Lie
\cite{Lie1,Lie2} and completed by his student A. Tresse for
second order differential equations (\ref{1}) with arbitrary
smooth functions $f(x,y,y')$ in \cite{Tresse}. The geometrical
description of these results was given by E. Cartan
\cite{Cartan1,Cartan2}, when he introduced the new idea of
projective connections.  It was important for the theory of
invariants to investigate the invariants themselves and their
possible connections among each other, because it enables us to
say whether our equation (\ref{1}) allows some infinitesimal group
of point transformations or not
\cite{Tresse}. In projective geometry it is important to know if
our invariant is equal to zero or not, because this circumstance
is responsible for the existence of corresponding geometrical
properties.

We will use these ideas to investigate the nonlinear dynamical
system of the type
\begin{eqnarray}
\dot{x}=P(x,y,z)\,,\nonumber\\
\dot{y}=Q(x,y,z)\,,\nonumber\\
\dot{z}=R(x,y,z) \,.\label{3}\end{eqnarray}
At first we conduct some ``projections'' of this system on, for
instance, the $(x,y)$-plane or the $(y,z)$-plane or some
nonplanar surface. After that we deal with some ordinary
differential equations.

In all our investigated cases these equations have the form
\begin{equation}y''+a_1(x,y)y'^3+3a_2(x,y)y'^2+3a_3(x,y)y'+a_4(x,y)=0\,.
\label{4}\end{equation}
It is well known that this form will be keeped under the point
transformations and invariants of such equations can be easy
investigated.  Hereforth we can look for conditions for the
parameters in (\ref{3}) in which one or more invariants are equal
to zero. We hope that exactly these conditions will be able to
show us where integrable cases lie.    It
is well known that some typical Riemann surfaces cannot yield a
satisfactory corresponding geometrical interpretation of the
equation (\ref{4}) in general, because metrical properties are
noninvariant under the point transformations. The adequate
geometrical image is only possible with the help of the notion
of normal projective connection.

According to E. Cartan \cite{Cartan1,Cartan2} we will use the following 
notations. 
For each second order differential equation we will look at the 
ccorresponding three dimensional vareity $M^3(x,y,y')$ of elements of 
first order with local coordinates $(x,y,y')$ and projective connections. 
The structure of this variety can be described through the 1-form 
components of projective connection
\[w^i,~w^0_i,~w^i_i-w_0^0,~w^j_i,~~i\ne j,~i,j=1,2,3\]
and 2-forms of curvature and torsion
\[\Omega^i,~\Omega_i^0,~\Omega_i^i-\Omega_0^0,~\Omega_i^j\,.\]
However, in the case of equations of the type (\ref{4}) a torsion free 
variety with normal projective connection is produced and our variety $M^3(
x,y,y')$ can be represented as a direct product
\[M^3(x,y,y')=V^2(x,y)\times S^1\,.\]
The 2-forms of curvature and torsion satisfy the following conditions
\begin{equation}\begin{array}{l}
\Omega^1=\Omega^2=0,~~\Omega^2_1=\Omega^1_2=0,~~\Omega_1^1-\Omega_0^0=
\Omega_2^2-\Omega_0^0=0,\nonumber\\
\Omega_1^0=b~\omega^1\wedge\omega^2,~~\Omega_2^0=h~\omega^1\wedge\omega^2\,,\
\label{5}\end{array}\end{equation}
where
\begin{eqnarray}
b&=&
2{\partial^2a_2\over\partial x\partial y}-
{\partial^2a_3\over\partial y^2}-
{\partial^2a_1\over\partial x^2}+
2a_1{\partial a_4\over \partial y}+
a_4{\partial a_1\over \partial y}-
3a_1{\partial a_3\over \partial x}-
3a_3{\partial a_1\over \partial x}- \nonumber \\
&&3a_2{\partial a_3\over \partial y}+
6a_2{\partial a_2\over \partial x}+
\left( 2{\partial^2a_3\over\partial x\partial y}-
{\partial^2a_2\over\partial x^2}-
{\partial^2a_4\over\partial y^2}-
2a_4{\partial a_1\over \partial x}-\right.\nonumber \\ 
&&\left. a_1{\partial a_4\over \partial x}+
3a_4{\partial a_2\over \partial y}+
3a_2{\partial a_4\over \partial y}+
3a_3{\partial a_2\over \partial x}-
6a_3{\partial a_3\over \partial y}\right)~y', \label{6}\\
h&\equiv&{\partial b\over \partial y'}\,.\nonumber \end{eqnarray}

The components of the normal projective connections also have a very 
simple form
$$\begin{array}{l}
w^1={\rm d}x,~w^2={\rm d}y,~w_1^2=a_4{\rm d}x+a_3{\rm d}y,~w_2^1=-a_2{\rm 
d}x-a_1{\rm d}y,\\
w_1^1=-w_2^2=-a_3{\rm d}x-a_2{\rm d}y,~w_1^0=\Pi^0_{11}{\rm d}x+\Pi^0_{
12}{\rm d}y,~w_2^0=\Pi^0_{21}{\rm d}x+\Pi^0_{22}{\rm d}y,
\end{array}$$
with
$$\begin{array}{l}
\Pi^0_{11}=2(a_3^2-a_2a_4)+a_{3x}-a_{4y}\,,\\
\Pi^0_{22}=2(a_2^2-a_1a_3)+a_{1x}-a_{2y}\,,\\
\Pi^0_{12}=\Pi^0_{21}=a_2a_3-a_1a_4+a_{2x}-a_{3y}\,.
\end{array}$$
The Cartan's structure equations have the form
$$\begin{array}{l}
{\rm d}w^i=w^k\wedge w^i_k\,,\\
{\rm d}w^j_i=w_i^0\wedge w^j+w^k_i\wedge w^j_k-\delta_i^jw^k\wedge 
w^0_k\,,\\
{\rm d}w^0_i=w^k\wedge w^0_k-\frac{1}{2}R^0_{ikl}w^k\wedge w^l~,
 i,j,k,l=1,2\,.
\end{array}$$
The tensor of projective curvature has two components $R^0_{112}$ and $R^
0_{212}$, that will be noted as $-L_1$ and $-L_2$ respectively:
\begin{eqnarray}
L_1=-{\partial\Pi^0_{11}\over\partial y}+
{\partial\Pi^0_{12}\over\partial x}-a_2\Pi^0_{11}-a_4\Pi^0_{22}+2a_3\Pi^0_{
12}\,,\nonumber \\
L_2=-{\partial\Pi^0_{12}\over\partial y}+
{\partial\Pi^0_{22}\over\partial x}-a_1\Pi^0_{11}-a_3\Pi^0_{22}+2a_2\Pi^0_{
12}\,.\label{7}\end{eqnarray}

Both these values were discovered by R. Liouville much earlier than by E. 
Cartan and used for the construction of the semi invariant $\nu_5$ 
with the weight equal to 5.
This invariant is a semi invariant. This means that after some point 
transformation (\ref{2}) the new $\tilde{\nu_5}$ is equal to
\[\tilde{\nu_5}=\Delta^5\nu_5\,,\]
where $\Delta\ne 0$ is the functional determinant of our transformation
in opposite to an absolute invariant where we have $\tilde{\tau}=\tau$. 
The degree of this determinant is exactly the weight of the corresponding 
invariant. The value of $\nu_5$ can be calculated by means of the 
following formula:
\begin{equation}
\nu_5=L_2(L_1L_{2x}-L_2L_{1x})+L_1(L_2L_{1y}-L_1L_{2y})-a_1L_1^3+3a_2L_1^
2L_2-3a_3L_1L_2^2+a_4L_2^3\,.\label{8}\end{equation}

If $\nu_5\ne 0$ we can use the recursive formula of R. Liouville 
\cite{Liouville} 
to build the series of semi invariants with higher wights:
\begin{equation}
\nu_{m+2}=L_1{\partial\nu_m\over\partial y}-L_2{\partial\nu_m\over\partial 
x}+m\nu_m(L_{2x}-L_{1y}),~~m\ge 5\,,\label{9}\end{equation}
and after that construct the absolute invariants
\begin{equation}t_m=\nu_m\nu_5^{-m/5}\,.\label{10}\end{equation}
In the case $\nu_5=0$, R. Liouville has found other seires of semi 
invariants. At first, he found some invariants of weight 1:
\[
w_1={1\over L_2^4}\left[L_2^3(\alpha 'L_2-\alpha L_1)-R_2(
L_2^2)_y+L_2R_{2y}-L_2R_2(a_1L_1-a_2L_2)\right]
\]
for $L_2\ne 0$, and
\begin{equation}
w_1={1\over L_1^4}\left[L_1^3(\alpha 'L_1-\alpha '' L_2)+R_1(
L_1^2)_x-L_1^2R_{1x}+L_1R_1(a_3L_1-a_4L_2)\right]
\label{11a}\end{equation}
for $L_1\ne 0$, where
\begin{eqnarray}
R_1&=&L_1L_{2x}-L_2L_{1x}+a_2L_1^2-2a_3L_1L_2+a_4L_2^2,\nonumber\\
R_2&=&L_1L_{2y}-L_2L_{1y}+a_1L_1^2-2a_2L_1L_2+a_3L_2^2,\nonumber \\
\alpha&=&a_{2y}-a_{1x}+2(a_1a_3-a_2^2),~\alpha '=a_{3y}-a_{2x}+a_1a_4-a_2a_3,
\nonumber\\
\alpha ''&=&a_{4y}-a_{3x}+2(a_2a_4-a_3^2)\,. \label{12}\end{eqnarray}

Also in this case there exists one recursive formula to construct the 
series of semi invariants with higher weights
\begin{equation}
w_{m+2}=L_1{\partial w_m\over\partial y}-L_2{\partial w_m\over\partial x}+
mw_m(L_{2x}-L_{1y})
\label{13}\end{equation}
and the corresponding absolute invariants read
\begin{equation}
u_{m+2}={w_{m+2}\over w_1^{m+2}},~~m\ge 1\,.\label{14}\end{equation}
The case that both projective curvatures $L_1$ and $L_2$ are equal to zero 
is non interesting, while the equation (\ref{4}) can then be reduced to $
y''=0$ \cite{Liouville}.
If also $w_1=0$ there are other series of invariants. The first one has 
weight equal to 2 and is given by
\begin{equation}
i_2={3R_1\over L_1}+{\partial L_2\over\partial x}-{\partial L_1\over
\partial y}
\label{15}\end{equation}
and other semi invariants of this series by
\begin{equation}
i_{2m+2}=L_1{\partial i_{2m}\over\partial y}-L_2{\partial i_{2m}\over
\partial x}+2mi_{2m}\left({\partial L_2\over\partial x}-{\partial L_1\over
\partial y}\right)\,,~(m\ge 1).\label{16}\end{equation}
 The corresponding absolute invariants can be found through the relation
\begin{equation}j_{2m}=i_{2m}i_2^{-m}\,.\label{17}\end{equation}
If for our equation (\ref{4}) the semi invariant
 
\begin{equation} \nu_5=0 \label{18a}\end{equation}

and $L_1,L_2\ne 0$ that we can immediately construct the first integral of 
this equation
\begin{equation}
y'=-{L_1\over L_2}\,.\label{18}\end{equation}
For many physical systems like (\ref{3}) those partial solutions are very 
interesting.

On the other hand side the condition (\ref{18a}) has also a deep geometrical 
sense. Namely, if and only if the condition (\ref{18a}) is fulfilled it will 
be possible to immerse our variety $V^2(x,y)$ of the normal projective 
connection into the real projective space ${\Bbb P}^3({\Bbb R})$. In all 
other cases the immersion will be possible only in ${\Bbb P}^4({\Bbb R})$
\cite{Dryuma85}.

The immersion into ${\Bbb P}^3({\Bbb R})$ will be realized on one of the 
simplest surfaces - on developed surfaces. It is well known that the 
coresponding surfaces for the equation $y''=0$ is the projective plane. In 
this sense the equation (\ref{4}) with the condition (\ref{18a}) are really 
the next simple ones. As we mentioned above, the condition that some 
invariant is equal to zero must have some geometrical sense, is in 
the cases $w_1=0$ and $i_2=0$ not known to us.

The first investigation of dynamical systems by the described method was  
done in \cite{Dryuma,Dryuma85}. In the present work we deal with 
different aspects of the Lorenz system as well as with the R\"o\ss ler 
system.

\section{The Lorenz system}
In the work \cite{Lorenz} Lorenz discussed the problem of the representation 
of a forced dissipative hydrodynamic flow and suggested the following 
model for the description of that system:
\begin{eqnarray}
{{\rm d}x\over  {\rm d}t}&=&\sigma(y-x)\,,\nonumber\\
{{\rm d}y\over  {\rm d}t}&=&rx-y-zx\,,\nonumber\\
{{\rm d}z\over  {\rm d}t}&=&xy-bz\,,\label{18'}\end{eqnarray}
where $\sigma,r,b$ are some parameters characterizing the flow.
Now, this system is well investigated by some analytical and numerical 
methods.
It posseses several dynamical states corresponding to different regions in 
the parameter space. However, up to now no one analytical criterion is 
known which allows to decide wheter the for a given set of parameters the 
solution will be a regular or a stochastic one. 

The phase space of (\ref{18'}) is three dimensional and it is separeted 
due to the integral curves of the following two equations
\begin{equation}
{{\rm d}y\over {\rm d}x}={rx-y-xz\over\sigma (y-x)},~~
{{\rm d}z\over {\rm d}x}={xy-bz\over\sigma (y-z)}\,.
\label{19}\end{equation}
After eliminating one of the functions, $z$, for instance, and introducing 
new notations we get an equivalent second order differential equation of 
the type (\ref{4})
\begin{equation}
{{\rm d}^2y\over{\rm d}x^2}-{3\over y}{{\rm d}y\over{\rm d}x}^2+
\left(\alpha y-{1\over x}\right){{\rm d}y\over{\rm d}x}+
\epsilon xy^4+{\delta y^2\over x}-\gamma y^3-\beta x^3y^4-\beta x^2y^3=0
\label{20}\end{equation}
with
\begin{equation}
\alpha={b+\sigma+1\over \sigma},~
\beta={1\over \sigma^2},~
\gamma ={b(\sigma +1)\over \sigma^2},~
\delta={\sigma+1\over\sigma},~\epsilon ={b(r-1)\over \sigma^2}\,.
\label{21}\end{equation}
Roughly speaking, this is the projection of the system (\ref{18'}) on 
the $(x,y)$-plane. The equation (\ref{20}) was investigated in the paper 
\cite{Dryuma}. The semi invariant $\nu_5$ has the form
\begin{equation}
\nu_5 =\tilde{u}_1z+\tilde{u}_2t^2+\tilde{u}_4\label{22}\end{equation}
with $z=x^2$, $t=1/(xy)$ and
\begin{eqnarray}
\tilde{u}_1&=&\alpha\beta (10\alpha -\alpha^2-6\delta ),\nonumber \\
\tilde{u}_2&=&\frac{2}{9}\alpha (\alpha +3\delta )(2\alpha -3\delta ),
\nonumber \\
\tilde{u}_4&=&\alpha ({\alpha^2\over 9}(2\alpha^2-9\gamma )-2\epsilon (2\alpha 
-3\delta))\,. \label{23}\end{eqnarray}

The projective curvatures are
\begin{eqnarray}
L_1&=&3(\epsilon x-\beta x^3)y^2-\frac{2}{3}\alpha^2y+{1\over 3x}
(2\alpha-3\delta)\,,\nonumber \\
L_2&=&{\alpha\over y}\,.\label{24}\end{eqnarray}

The invariant $\nu_5$ is equal to zero in two cases. First, for $\alpha =0$, 
i.e., $\sigma=-b-1$. This case is noninteresting because it cannot be reached 
for 
physical values of the parameters. In the second case we have $\nu_5=0$ 
for arbitrary $r$ and $\sigma=2b-1$ with $b=0$ and $b=2/3$. This is a 
well known case of regular behaviour of the system (\ref{18'}). 

In the following we shall continue the investigation of \cite{Dryuma} and 
calculate the other invariants as well as the invariants of other 
projections of the system (\ref{18'}).

R. Liouville \cite{Liouville} found that in the case $\nu_5 =0$ and $L_1,
L_2\ne 0$ we can construct the first integral of the equation (\ref{4}):
\begin{equation}y'=-{L_1\over L_2}\,.\label{24a}\end{equation}
Consequently, for the projection on the $(x,y)$-plane with the parameters
\begin{equation}
\sigma=-1/5, ~b=-16/5, ~r=-7/5\label{27a}\end{equation} 
we get the equation
\begin{equation}y'=-{y\over x}(4-32xy+192x^2y^2-25x^4y^2)\,,
\label{24b}\end{equation}
where $y(x)$ is the solution of equation (\ref{20}). Let $xy=s$, then we get 
Abel's equation of the first kind \cite{Kamke}
\begin{equation}s_x=f_2(x)s^2+f_3(x)s^3\label{28a}\end{equation}
and after the substitution
\begin{eqnarray}
s(x)&=&w(x)\eta(\xi)-{f_2(x)\over 3f_3(x)},~
f_2(x)={8\over x},~
f_3(x)=-{42\over x}-{25\over 4}x,\nonumber\\
w(x)&=&c\left({x^2\over 168+25x}\right)^\frac{16}{63},\label{24d}\\
\xi&=&\int f_3(x)w^2(x){\rm d}x=
-{1\over 8}\left(168+25x^2\right)^\frac{31}{63}x^\frac{64}{63}-
{62\over 3}\int x^\frac{1}{63}(168+25x^2)^{-\frac{32}{63}}
{\rm d}x
\nonumber\end{eqnarray}
we have the following equation for the function $\eta(\xi)$
\[\eta'(\xi)=\eta^3(\xi)+I(x)\,,\]
with
\[ I(x)=-{256\over 27}{256+225x^2\over (168+25x^2)^{47/21}}\]
and $\xi$ as in equation (\ref{24d}).

We believe that this is a new first order integral of the Lorenz 
system, 
although it may be of purely mathematical interest due to the choice of 
parameters given by equation (\ref{27a}).

Another case when a first integral of type (\ref{24a}) exists corresponds 
to the regular behaviour of the Lorenz system, that is to 
\begin{eqnarray*}
\sigma=1/3,~b=2/3, ~~r~{\rm arbitrary}.\end{eqnarray*}
 Then we have
\[y'=y^2(4+3(1-r)xy+\frac{9}{2}x^3y)\,.\]
This is the well known Abel equation (\ref{28a})
with
$$
f_3=3(1-r)x+\frac{9}{2}x^3 ,~~f_2=4 $$
and after a substitution in analogy with (\ref{24d}) we get the canonical 
expression
\begin{eqnarray*}
y(x)&=&w(x)\eta (\xi)-{4\over 3(3(1-r)x+\frac{9}{2}x^3)},\\
w&=&\exp \left(\int {-16\over 3(3(1-r)x+\frac{9}{2}x^3)} {\rm d}x\right) 
\end{eqnarray*}
with
$$ (3(1-r)x+\frac{9}{2}x^3)w^3I(x)={{\rm d}\over {\rm d}x}
\left({4\over  3(3(1-r)x+\frac{9}{2}x^3)}  \right)+
{128\over 27} (3(1-r)x+\frac{9}{2}x^3) $$
and
\[\eta'(\xi)=\eta^3+I(x)\,.\]
To our knowledge, there are no such solutions in the literature.

Let $\nu_5\ne 0$, then our two dimensional variety $V_2(x,y)$ can be 
immersed into ${\Bbb P}^4(\Bbb{R})$ only. On the other hand side we can 
calculate the invariants of higher weights and look for conditions on the 
parameters that one  of that invariants vanish or all invariants are 
functions of one of them. In the first case we can hope that our immersed 
surface has some special geometrical properties. In the second case we 
can decide wheter our differential equation (\ref{20}) possesses some 
group of infinitesimal transformations or not.

We calculated $\nu_7$, $\nu_9$, $\nu_{11}$,  and some resultants of $\nu_5$ 
and $\nu_7-R_{57}$, $\nu_5$ and $\nu_9-R_{59}$, $\nu_7$ and $\nu_9-R_{
79}$, as well as the resultants of $R_{57}$, $R_{59}$, $R_{79}$. The 
results of these calculations are quite voluminous and only some first of 
them will be listed here. This gives us enought insight into the structure 
of them.

The semi invariant of weight $7$ for (\ref{20}) equals to
\begin{equation}\nu_7=(v_1z^2+v_2zt^2+v_3zt+v_4z+v_5t^4+v_6t^3+v_7t^2+
v_9t+v_{10})/t\label{25}\end{equation}
with $z=x^2,~t=1/(xy)$ and
\begin{eqnarray}
v_1&=&30\beta \tilde{u}_1,~
v_2=2\alpha \beta (\alpha -2)(\alpha^2+18\delta),~
v_3=\frac{10}{3}\alpha^2\tilde{u}_1,~\nonumber\\ 
v_4&=&10\alpha\beta ({\alpha^2\over 3}(2\alpha^2-9\gamma)+
		3\epsilon(\alpha^2-14\alpha+12\delta)),~
v_5=\frac{2}{9}(\alpha+3\delta)\tilde{u}_2,\label{27}\\
v_6&=&\frac{28}{27}\alpha^3(\alpha+3\delta)(2\alpha-3\delta ),~
v_7=-\frac{54}{7}\epsilon\alpha^{-2}v_6,~
v_9=\frac{10}{3}\alpha^2\tilde{u}_4,~
v_{10}=30\epsilon \tilde{u}_4\,\nonumber \end{eqnarray}
with $\tilde{u}_1,\tilde{u}_2,\tilde{u}_4$ defined in (\ref{23}).
For the semi invariant of weight 9 we found the representation
\begin{eqnarray}
\nu_9 & = & (\tilde{w}_1z^3+\tilde{w}_2z^2t^2+\tilde{w}_3z^2t+\tilde{w}_4z^2+
\tilde{w}_5zt^4+
\tilde{w}_6zt^3+
\tilde{w}_7zt^2+
\tilde{w}_8zt+\nonumber\\
& & \tilde{w}_9z+\tilde{w}_{10}t^6+\tilde{w}_{11}t^5+\tilde{w}_{12}t^4+\tilde{
w}_{13}t^3+\tilde{w}_{14}t^2+\tilde{w}_{16}t+\tilde{w}_{17})
/t^2\,,\label{28}\end{eqnarray}
with $z$ and $t$ as above and
\begin{eqnarray}
\tilde{w}_1 & = & 1170\beta^2\tilde{u}_1,~
\tilde{w}_2=10\alpha\beta^2(-148\alpha^2+22\alpha^3+210\alpha+\delta
	+3\alpha^2\delta -306\delta ^2),~ \nonumber\\
\tilde{w}_3 & = & 260\alpha^2\beta \tilde{u}_1,~
\tilde{w}_4=130\alpha\beta^2(\alpha^2(2\alpha^2-9\gamma
	)+18\epsilon(\alpha^2-12\alpha+9\delta )),~\nonumber\\
\tilde{w}_5 & = & \frac{2}{9}\alpha^4\beta(98-15\alpha)+2\alpha
\beta((\alpha^2\
delta
	(2+\delta )+\delta ^2(78\alpha-138\delta )),~\nonumber\\
\tilde{w}_6&=&52\alpha^2\beta(\alpha^2(2+\alpha)+6\delta
	(11\alpha-24\delta ))/9,~\nonumber \\
\tilde{w}_7& = &10\alpha\beta(-\alpha^4(\frac{14}{9}\alpha^2-14\alpha+10
\delta
	)+4\alpha^2\epsilon(11-4\alpha)+\nonumber\\ 
	& & 3\delta \epsilon(204
	\delta^2-104\alpha-\alpha^2)+
	\alpha^2\gamma (7\alpha+3\delta )),~\nonumber\\
\tilde{w}_8 & = & 260\alpha^2\beta(\frac{1}{9}\alpha^2(2\alpha^2-9\gamma
	)+9\epsilon(\alpha^2-14\alpha+12\delta )),~\label{29} \\
\tilde{w}_9 & = & -130\alpha\beta\epsilon(2\alpha^2(2\alpha^2-9\gamma
	)+9\epsilon(\alpha^2- 18\alpha+18\delta )),~\nonumber\\
\tilde{w}_{10} & = & \frac{2}{9}(\alpha+3\delta )^2\tilde{u}_2,~
\tilde{w}_{11}=\frac{68}{27}\alpha^2(\alpha+3\delta )\tilde{u}_2,~
\tilde{w}_{12}=2(14\alpha^4-23\epsilon(\alpha+3\delta ))\tilde{u}_2,~
\nonumber\\
\tilde{w}_{13} & = & -416\epsilon\alpha^2\tilde{u}_2,~\nonumber\\
\tilde{w}_{14} & = & 10\alpha(\frac{14}{81}\alpha^6(2\alpha^2-9\gamma
	)-2\epsilon\alpha^4(3\alpha-5\delta
	)+34\epsilon^2(2\alpha^2+3\alpha\delta -9\delta ^2)-\nonumber\\&&~~~~
	\alpha^2\epsilon\gamma (\alpha+3\delta )),~ \nonumber\\
\tilde{w}_{16} & = & -260\alpha^2\epsilon \tilde{u}_4,~ 
\tilde{w}_{17}=1170 \epsilon^4 \tilde{u}_4\,
\nonumber\end{eqnarray}
with $\tilde{u}_1,\tilde{u}_2,\tilde{u}_4$ from (\ref{23}).

The first two of the resultants $R_{57}$ and $R_{59}$ have a short form. 
Using the notions from (\ref{23}) and (\ref{27}-\ref{29}) we have
\begin{eqnarray*}
R_{57}&=&\nu_5^2\tilde{u}_3^2v_1+\tilde{u}_4^2v_1+\tilde{u}_1^2v_{10}-
\tilde{u}_1\tilde{u}_4v_4+
\nu_5(2\tilde{u}_3\tilde{u}_4v_1-\tilde{u}_1\tilde{u}_3v_4)+
	t^4(\tilde{u}_2^2v_1-\tilde{u}_1\tilde{u}_2v_2+        \\
&&\tilde{u}_1^2v_5)
	+t^3(\tilde{u}_1^2v_6-\tilde{u}_1\tilde{u}_2v_3)+
t^2(2\tilde{u}_2\tilde{u}_4v_1-
\tilde{u}_1\tilde{u}_4v_2)+
	\nu_5(2\tilde{u}_2\tilde{u}_3v_1-    \\
&&\tilde{u}_1\tilde{u}_3v_2)-
\tilde{u}_1\tilde{u}_2v_4+\tilde{u}_1^
2v_7+t(\nu_7\tilde{u}_1^2v_8+\tilde{u}_1^2v_9-\nu_5\tilde{u}
_1\tilde{u}_3v_3-
\tilde{u}_1\tilde{u}_4v_3),
\end{eqnarray*}

\begin{eqnarray*}
R_{59}&=&\nu_5^3 \tilde{u}_3^3 \tilde{w}_1 + \tilde{u}_4^3 \tilde{w}_1 - 
\tilde{u}_1^3 \tilde{w}_17 - \tilde{u}_1 \tilde{u}_4^2 \tilde{w}_4 + 
  \nu_5^2 (3 \tilde{u}_3^2 \tilde{u}_4 \tilde{w}_1 - \tilde{u}_1 
\tilde{u}_3^2 \tilde{w}_4) + 
  t^6 (\tilde{u}_2^3 \tilde{w}_1 -             \\&&
\tilde{u}_1^3 \tilde{w}_{10} - 
\tilde{u}_1 \tilde{u}_2^2 \tilde{w}_2 + 
\tilde{u}_1^2 \tilde{u}_2 \tilde{w}_5) +  
t^5 (- \tilde{u}_1^3 \tilde{w}_{11} - 
\tilde{u}_1 \tilde{u}_2^2 \tilde{w}_3 + 
\tilde{u}_1^2 \tilde{u}_2 \tilde{w}_6) +     \\&&
  t^4 (3 \tilde{u}_2^2 \tilde{u}_4 \tilde{w}_1 - 
\tilde{u}_1^3 \tilde{w}_{12} - 
2 \tilde{u}_1 \tilde{u}_2 \tilde{u}_4 \tilde{w}_2 - 
\tilde{u}_1 \tilde{u}_2^2 \tilde{w}_4 + 
     \tilde{u}_1^2 \tilde{u}_4 \tilde{w}_5 + 
\nu_5 (3 \tilde{u}_2^2 \tilde{u}_3 \tilde{w}_1 - \\&&
2 \tilde{u}_1 \tilde{u}_2 \tilde{u}_3 \tilde{w}_2 + 
        \tilde{u}_1^2 \tilde{u}_3 \tilde{w}_5) + 
\tilde{u}_1^2 \tilde{u}_2 \tilde{w}_7) + 
  t^3 (- \tilde{u}_1^3 \tilde{w}_{13} - 
2 \tilde{u}_1 \tilde{u}_2 \tilde{u}_4 \tilde{w}_3 + 
\tilde{u}_1^2 \tilde{u}_4 \tilde{w}_6 + \\&&
     \nu_5 (-2 \tilde{u}_1 \tilde{u}_2 \tilde{u}_3 \tilde{w}_3 + 
\tilde{u}_1^2 \tilde{u}_3 \tilde{w}_6) + 
\tilde{u}_1^2 \tilde{u}_2 \tilde{w}_8) + 
  t (- \tilde{u}_1^3 \tilde{w}_{16} - 
\nu_5^2 \tilde{u}_1 \tilde{u}_3^2 \tilde{w}_3 - \\&&
\tilde{u}_1 \tilde{u}_4^2 \tilde{w}_3 + 
\tilde{u}_1^2 \tilde{u}_4 \tilde{w}_8 + 
     \nu_5 (-2 \tilde{u}_1 \tilde{u}_3 \tilde{u}_4 \tilde{w}_3 + 
\tilde{u}_1^2 \tilde{u}_3 \tilde{w}_8)) + 
\tilde{u}_1^2 \tilde{u}_4 \tilde{w}_9 + \\&&
  t^2 (3 \tilde{u}_2 \tilde{u}_4^2 \tilde{w}_1 - 
\tilde{u}_1^3 \tilde{w}_14 - 
\nu_9 \tilde{u}_1^3 \tilde{w}_15 - 
\tilde{u}_1 \tilde{u}_4^2 \tilde{w}_2 + 
     \nu_5^2 (3 \tilde{u}_2 \tilde{u}_3^2 \tilde{w}_1 - 
\tilde{u}_1 \tilde{u}_3^2 \tilde{w}_2) - \\&&
2 \tilde{u}_1 \tilde{u}_2 \tilde{u}_4 \tilde{w}_4 + 
     \tilde{u}_1^2 \tilde{u}_4 \tilde{w}_7 + 
\nu_5 (6 \tilde{u}_2 \tilde{u}_3 \tilde{u}_4 \tilde{w}_1 - 
2 \tilde{u}_1 \tilde{u}_3 \tilde{u}_4 \tilde{w}_2 - 
        2 \tilde{u}_1 \tilde{u}_2 \tilde{u}_3 \tilde{w}_4 + \\&&
\tilde{u}_1^2 \tilde{u}_3 \tilde{w}_7) + 
\tilde{u}_1^2 \tilde{u}_2 \tilde{w}_9) + 
  \nu_5 (3 \tilde{u}_3 \tilde{u}_4^2 \tilde{w}_1 - 
2 \tilde{u}_1 \tilde{u}_3 \tilde{u}_4 \tilde{w}_4 + 
\tilde{u}_1^2 \tilde{u}_3 \tilde{w}_9)\,.
\end{eqnarray*}

Based on the analysis of the formulas (\ref{25},\ref{29}) we raise the 
supposition that each of the following higher invariants $\nu_7$, $\nu_9$, 
$\nu_{11}$ can be equal 
to zero for the same values of the 
parameters as $\nu_5$. We failed to find different conditions on the 
parameters of the system (\ref{18'}). Because all semi  invariants of this 
series can be found by use of the recusive relation (\ref{9}) this 
supposition will we certainly true for all of them. Exactly the same can be 
said 
about the structure of the resultants. Let us remark that all these 
formulas become considerably shorter and simpler in the case $\sigma=2b-
1$ and only in this case we obtained a transparent relation between the 
invariants.Really we see,that for $\sigma=2b-1$ 
\begin{eqnarray}
\nu_5  &=&  {27 b^2 (-2 + 3 b)z\over (2 b-1 )^5}\,,\nonumber\\ 
\nu_7 & =&  {162 b^2 (-2 + 3 b) z (5b - 5bq + 5b^2t + bt^2 - 2b^2t^2+5z)\over
(2b-1)^7}\,,\nonumber \\ 
\nu_9  &=&  {486b^2(-2 + 3b)z\over ( 2b-1 )^9}
\left(65b^2(1 - q)^2 + 130b^3t(1 - q) +
\right. \nonumber\\
 & & \left.
10 b^2t^2( 3(1- 2b)(1-q) + 7b^2 ) + 26b^3t^3(1 - 2b) + \right. \nonumber\\
 & & \left.
b^2t^4(1 - 2b)^2 + 
130bz(1 - q) +130b^2tz + 40bt^2z(1 - 2b) + 65z^2)\right)\,,\nonumber\\
R_{57}&=&{81(2 -3b)b^2 \over(2b-1)^{12}} (10\nu_5^2    (-1 + 2 b)^5+ 
54 \nu_5 b^3(2- 3b)(-5(1-q) -\nonumber\\&&5 bt-(1-2b)t^2)+  
9\nu_7 (2 - 3b)b^2(1 - 2b)^2 t))\,.
\label{125}\end{eqnarray}

 Also, in this case we have a good correspondence with the 
physical interpretation of the system. The case $\sigma=2b-1$ is the well 
known integrabel case and the system does not possess any stochastic 
behaviour. 

The investigated projection at the $(x,y)$-plane was the simplest one. In 
the following we shall take more complicated projections and look for 
conditions on the parameters.
In the first out from the more complicated cases we start with the 
substitution $\xi=x^2-bz$ in the system (\ref{19}) and eliminate the 
variable $y$ from both equations. In this way we obtain the $(x,\xi )$-
projection. For convenience we shall use the notation $y$ instead of 
$\xi$. The second order differential equation which we obtain in this way 
has 
the canonical 
form given by equation (\ref{4}) with functions $a_i(x,y),~i=1,...,4$ as 
follows:
\begin{eqnarray}
a_1(x,y)&=&{\sigma (r-1)x\over b^2 y^2}+{\sigma x\over b^3y}-
{\sigma x^3\over b^3y^2}\,,\nonumber\\
a_2(x,y)&=&{\sigma +1\over 3by}-{(2\sigma-b)(r-1)x^2\over b^2y^2}-
{(2\sigma -b)x^2\over b^3y}+{(2\sigma -b)x^4\over b^3y^2}-{1\over 3 y}\,,
\nonumber\\
a_3(x,y)&=&{(r-3)(2\sigma-b)^2x^3\over b^2\sigma y^2}-{2(2\sigma -b)(\sigma+1)
x\over 3 b \sigma y}+\nonumber\\
&&{(2\sigma -b)^2x^3\over b^3\sigma y}-{(2\sigma -b)^2x^5\over b^3\sigma y^
2}+{(2\sigma -b)x\over 3\sigma y}\,,\label{366}\\ 
a_4(x,y)&=&{(2\sigma -b)^3x^6\over b^3\sigma^2 y^2}-{(2\sigma-b)^3x^4\over 
b^3\sigma^2y}+{(\sigma+1)(2\sigma -b)^2x^2\over \sigma^2by}-\nonumber\\
&&{(r-1)(2\sigma -b)^3x^4\over \sigma^2b^2y^2}-{2\sigma -b\over \sigma}\,.
\nonumber\end{eqnarray}
In this projection the projective curvatures have the form
\begin{eqnarray*}
L_1&=&{2\sigma -b\over 3b^2\sigma^2y^4}
	\left( 9x^4b(b-2\sigma)^2(1-r)+9x^6(b-2\sigma)^2+2x^2yb(-b(1+b)^2+
\right.\\ 
&&~~~~~~   2\sigma(1+8b-6r(b-\sigma)-4\sigma+\sigma^2+3b\sigma))
	+30x^4y\sigma(b-2\sigma)-\nonumber\\&&~~~~~\left. y^2b^2\sigma
(1+b+\sigma )-
	3x^2y^2\sigma(b-8\sigma )\right)\,,\\
L_2&=&{x\over 3b^3\sigma y^4}
	\left[ -9x^2b(b-2\sigma)^2(1-r)-
	9x^4(b-2\sigma)^2+yb(2b+b^2-b^3+ \right.\\ 
&& 
~~~~~~ \sigma(-4-10b+	3b^2+16\sigma +
	4\sigma^2+12r(b-2\sigma)))-
	30x^2y\sigma (b-2\sigma)+\\&&~~~~~\left.
	3y^2\sigma (b-8\sigma)\right]\,,
\end{eqnarray*}
and the semi invariant $\nu_5$ is given by
\begin{eqnarray*}
\nu_5&=&{(1+b+\sigma)(2\sigma -b)\over 27b^7\sigma^5y^{10}}\Big[
3b^2(2\sigma -b)^4(-1+b+2b^2-5\sigma+b\sigma +3r\sigma-\sigma^2)\times \\ &&
~~~~~~	 (2+b-b^2-5\sigma+b\sigma+9r\sigma +2\sigma^2)x^5+  
9b(2\sigma -b)^4(3b+6 b^2+3b^3+\\&& 
~~~~~~	\sigma-34b\sigma- 17b^2\sigma+18br\sigma+
	20\sigma^2-19b\sigma^2-18r\sigma^2+\sigma^3)x^7+
 \\ &&
~~~~~~
	 81\sigma(\sigma -2b)(2\sigma -b)^4x^9+18b\sigma(2\sigma -b)^3
(-3b-6b^2-3b^3-\sigma+
34b\sigma+\\ &&
~~~~~~17b^2\sigma-
	18br\sigma -20\sigma^2+19b\sigma^2+18r\sigma^2-\sigma^3)x^5y+
 \\ &&
~~~~~~324(2b-\sigma )\sigma^2(2\sigma -b)^3x^7y+
	b^2\sigma^2(2\sigma -b)^2 
(-2-b+b^2+5\sigma-b\sigma-\\ &&
~~~~~~9r\sigma-2\sigma^2) 
(3+2b-b^2-10\sigma+2b\sigma+16r\sigma+3\sigma^2)xy^2+b\sigma^2
(2\sigma -b)^2\\ &&
~~~~~~(33b+48b^2+15b^3+59\sigma-206b\sigma-
	139b^2\sigma+
144br\sigma-170\sigma^2-95b\sigma^2+
	\\ &&
~~~~~~
	288r\sigma^2+59\sigma^3)x^3y^2+
 18\sigma^2(2\sigma -b)^2(3b^2-44b\sigma+10\sigma^2)x^5y^2+\\ &&
~~~~~~2b\sigma^3(2\sigma -b)(7b+2b^2-
5b^3-59\sigma-44b\sigma+39b^2\sigma+36br\sigma+
170\sigma^2\\&&~~~~~~-15b\sigma^2-288r\sigma^2-59\sigma^3)xy^3+
 36(b-8\sigma)(b-2\sigma )\sigma^3(b+2\sigma)x^3y^3-\\&&~~~~~~
	9\sigma^4(8\sigma-b)^2xy^4)~\Big ]\,.
\end{eqnarray*}
In this case we can succeed in $\nu_5$ beeing identically equal to zero for 
\begin{equation}
\sigma=-1-b ~~~{\rm and}~~~b=2\sigma
\label{33}\end{equation}
only. The first condition belongs to the mentioned above unphysical case, 
the second one represents the integrable case. The surface $V_2(x,y)$  
can be immersed as a developeable surface in the projective space ${\Bbb 
P}^3({\Bbb R})$.

For $b=2\sigma$ we can investigate other series of invariants. The 
following ones (see above (\ref{11a})-(\ref{13})) with the first weight 
equal 
to 1 will be absent too because of $w_1\equiv 0$ for the considered 
values of the parameters. The next series of semi invariants $i_{2k},k=1,
2,...$ (\ref{15})-(\ref{16}) reads
$$
i_2={3\over 4\sigma^2y^2},~~i_4={9\over 8\sigma^4y^4},~~
i_6={27\over 8\sigma^6y^6},~~i_8={243\over 16\sigma^8y^8},\,...\,.$$
The projective curvatures have a very simple form in this case
\[L_1=0,~~~L_2=-{3x\over 4\sigma^2y^2} \]

and the functions $a_i$ (\ref{366}) are given by
$$
a_1={x\over 8\sigma^2y^2}(y-x^2+2q\sigma-2\sigma),~~
a_2={1-\sigma\over 6\sigma y},~~a_3=a_4=0\,.
$$
The corresponding absolute invariants $j_{2k}$ (\ref{17}) are constants:
\[j_4=2,~~j_6=8,~~j_8=48,\,...\,.\]

The last of the investigated projections of the Lorenz system is the (y,z)-
projection. The coefficients of the ordinary differential equation (\ref{4})
 have in this case the following form:
\begin{eqnarray*}
a_1(x,y)&=&
{\sigma y^4 - b y^2 x - b \sigma y^2 x + b^2 r x^2 - b^2 x^3\over (-y^2 + b 
r 
x - b x^2)^2},\\ 
a_2(x,y)&=&
{y (b y^2 + \sigma y^2 - 3 r \sigma y^2 - b^2 r x + 2 b r \sigma x + 3 
\sigma y^2 x + 
      b^2 x^2 - 2 b \sigma x^2)\over 3 (-y^2 + b r x - b x^2)^2},\\ 
a_3(x,y)&=&
(r y^2 (1 - b - 2 \sigma + 3 r \sigma ) + b r^2 x (-1 + b - \sigma ) +  
 y^2 x ( 1 + 2 \sigma - 6 r \sigma ) + \\ &&~~~~~~ r x^2 ( 2 b - b^2 + 2 b 
\sigma )+ 
3 \sigma y^2 x^2 - b x^3 (1 + \sigma ))/
  (3 (-y^2 + b r x - b x^2)^2),\\ 
a_4(x,y)&=&
{y (r^2 \sigma - r^3 \sigma + y^2 - b r x - 2 r \sigma x + 3 r^2 \sigma x +
b x^2 + \sigma x^2 - 
      3 r \sigma x^2 + \sigma x^3)\over (-y^2 + b r x - b x^2)^2}.
\end{eqnarray*}

For convenience we substituted $z$ for $x$. This is one of the complicatest
cases and all formulas are quite large. Therefore we list only the projective 
curvatures
$L_1$ and $L_2$ her. The first projective curvature has the form:
\begin{eqnarray*}
L_1&=&
y ( r^2 y^2(4 - 3 b^2  + b^3  + 20 \sigma - 
      26 b \sigma + 8 b^2 \sigma - 12 r \sigma + 24 b r \sigma - 9 b^2
r \sigma + 4 \sigma^2 - \\ && ~~2 b \sigma^2 ) + y^4 (2 - 2 b^2 + 2
\sigma - 2 b \sigma - 3 r \sigma + 6 b r \sigma ) + r^3 x (- 4 b + 3
b^3 - b^4 +\\ && ~~ 16 b \sigma - 10 b^2 \sigma + b^3 \sigma - 24 b r
\sigma + 12 b^2 r \sigma - 4 b \sigma^2 + 2 b^2 \sigma^2 )-r y^2 x (8
+ 4 b +\\ && ~~ 10 b^2 - 2 b^3 - 40 \sigma +
 70 b \sigma -29 b^2 \sigma + 36 r \sigma - 96 b r \sigma + 39 b^2 r
\sigma -8 \sigma^2 + 6 b \sigma^2) + \\ && ~~ 3 \sigma y^4 x (1 - 2 b)
+ r^2 x^2 (12 b - 6 b^2 - 13 b^3 + 5 b^4 - 48 b \sigma + 46 b^2 \sigma
- 6 b^3 \sigma + \\ && ~~ 96 b r \sigma - 57 b^2 r \sigma + 12 b
\sigma^2 -8 b^2 \sigma^2 ) + 4 y^2 x^2 (1 - b - b^2 + b^3 + 5 \sigma -
11 b \sigma + \\ && ~~ 6 b^2\sigma - 9 r \sigma + 30 b r \sigma - 15
b^2 r \sigma + \sigma^2- b \sigma^2) + b r x^3( - 12 + 12 b + 14 b^2 -
10 b^3 + \\ && ~~48 \sigma -
 62 b \sigma + 15 b^2 \sigma - 144 r \sigma + 99 b r \sigma - 12
\sigma^2 + 10 b \sigma^2 + \sigma y^2 x^3 (12 - 48 b + \\ && ~~ 30 b^2
) +
 x^4 (4 b - 6 b^2 - 4 b^3 + 6 b^4 - 16 b \sigma + 26 b^2 \sigma - 10
b^3 \sigma + 96 b r \sigma - 75 b^2 r \sigma + \\ && ~~ 4 b \sigma^2 -
4 b^2 \sigma^2) +3 b \sigma x^5( -8 + 7 b ))/ (3 (-y^2 + b r x - b
x^2)^4), \end{eqnarray*} the second one

\begin{eqnarray*} L_2&=& r y^4 (6 - b - 5 b^2 + 2 b^3 - 10 \sigma + 15
b \sigma - 5 b^2 \sigma + 12 r \sigma - 24 b r \sigma + 9 b^2 r \sigma
-4 \sigma^2 + \\&& ~~ 2 b \sigma^2 ) +
 3 \sigma y^6 (1 - 2 b ) +2 b r^2 y^2 x (-2 + b + 2 b^2 - b^3 - 12
\sigma + 10 b \sigma - 2 b^2 \sigma + \\&& ~~ 12 r \sigma - 6 b r
\sigma + 2 \sigma^2 - b \sigma^2 ) +2 y^4 x (-3 + 2 b + 3 b^2 - 2 b^3
+ 5 \sigma - 13 b \sigma + 8 b^2 \sigma - \\ && ~~ 12 r \sigma + 36 b
r \sigma - 15 b^2 r \sigma + 2 \sigma^2 - 2 b \sigma^2 ) + b^2 r^3 x^2
(-2 - b + b^2 - 2 \sigma + b \sigma ) +\\&& ~~ b r y^2 x^2 (8 - 6 b -
8 b^2 +
 6 b^3 + 48 \sigma - 64 b \sigma +15 b^2 \sigma - 72 r \sigma + 45 b r
\sigma - 8 \sigma^2 + \\ && ~ ~ 6 \sigma y^4 x^2 (2 - 6 b + 5 b^2 ) +
b^2 r^2 x^3 (6 + 2 b - 4 b^2 + 6 \sigma - 4 b \sigma ) +2 b y^2 x^3
(-2 + 2 b + \\ && ~~ 2 b^2 - 2 b^3 - 12 \sigma + 22 b \sigma - 10 b^2
\sigma + 36 r \sigma - 17 b r \sigma + 2 \sigma^2 - 2 b \sigma^2 )+\\
&& ~~ b^2 r x^4 ( -6 - b + 5 b^2 - 6 \sigma + 5 b \sigma ) + b \sigma
y^2 x^4 ( -24 + 21 b ) + \\&& ~~ 2 b^2 x^5 (1 - b^2 + \sigma - b
\sigma ))/
  (3 (-y^2 + b r x - b x^2)^4).
\end{eqnarray*}
It is easy to observe that both curvatures are much simpler by $ \sigma=0 $
and are both equal to zero in case 
\begin{equation} \sigma=0,~b=-1, ~~r~{\rm arbitrary}.\nonumber \end{equation}
That means that the corresponding second order differential
equation, the $(y,z)$-projection of the system (\ref{19}),
are equivalent to the simplest equation $y''=0$ and can be
integrated. 

\section{The R\"o\ss ler System}

For the R\"o\ss ler system \cite{Haken}-\cite{Roessler76d}
\begin{eqnarray*}\dot{x}&=&-y-z,\nonumber\\
\dot{y}&=&x+ay\,,\nonumber\\ \dot{z}&=&b+xz-cz\,,\label{44}\end{eqnarray*}
$a,b,c$ being parameters,
we investigated  3 projections. In all of 
them we leaved the notion $y$ for the function and $x$ for the 
independent variable.  The $(y,z)$-projection of (\ref{44}) after a simple 
transformation the system can be given in the form of equations (\ref{4})
with coefficients $a_i$ as follows:
\begin{eqnarray}
&&a_1=0,~~a_2={1\over 3y},~~a_3=-{(x+2a)x-y\over 3xy},\nonumber\\
&&a_4={a^2x-b+ay+ax^2\over xy}-{(2a+c)x+ba+x^2\over y^2}\,.
 \label{41}\end{eqnarray}
The semi invariant $\nu_5$ has the value
\begin{eqnarray*}
\nu_5&=&{2a+x\over 9x^2y^{10}}(2ax^2(18a-2a^3-9b+9c)+\\
&&~~x^3(18a-8a^3+9b-27c)+
3a^2x^4+7ax^5+2x^6+10a^2y^2+9axy^2)\,.
\end{eqnarray*}
It is easy to see that the invariant $\nu_5$ cannot be identically equal to 
zero 
for 
any values of the constants. Consequently, for the projection (\ref{41}) 
there does not exist any immersion $V_2(x,y)$ into ${\Bbb P}^3(\Bbb R)$ and 
the 
two dimensional variety $V_2(x,y)$ can be immeresed into ${\Bbb P}^4(\Bbb R)$ 
only. Also, we cannot expect to find a first integral of the system
 (\ref{44}) in this way. We calculated also other invariants of this
series, namely $\nu_7$ with weight 7 and $\nu_9$ with \hbox{weight
9}. For shortness we represent here only the formula for $\nu_7$ (that
for $\nu_9$ is too large): \begin{eqnarray*} \nu_7&=&{1\over 27 x^3
y^{15}}(360 a^3 b (-18 a + 2 a^3 + 9 b - 9 c) x^3 + 360 a^2 (-36 a^2 +
4 a^4 +\\&& ~~ 5 a^3 b - 36 a c + 2 a^3 c + 18 b c - 9 c^2) x^4 + 90 a
(-216 a^2 + 48 a^4 + \\&& ~~ 18 a b + 2 a^3 b - 9 b^2 - 36 a c + 20
a^3 c + 27 b c + 36 c^2) x^5 +\\&& ~~
 90 (-108 a^2 + 24 a^4 - 18 a b - 17 a^3 b + 72 a c + 2 a^3 c - 9 b c
+ 27 c^2) x^6 +\\&& ~~ 90 (-18 a - 32 a^3 - 9 b - 11 a^2 b + 27 c - 17
a^2 c) x^7 - 90 a (39 a + 2 b + 11 c) x^8 +\\&& ~~ (-1350 a - 180 c)
x^9 - 180 x^{10} + 10 x^3 (2 a + x)^3 (36 a^2 - 4 a^4 -\\&& ~~ 18 a b
+ 18 a c + 18 a x - 8 a^3 x + 9 b x - 27 c x + 3 a^2 x^2 + 7 a x^3 +2
x^4) y + \\&& ~~
 27 x (2 a + x) (-40 a^3 b - 80 a^3 x - 36 a^2 b x - 40 a^2 c x - 124
a^2 x^2 + 10 a b x^2 -\\&& ~~ 50 a c x^2 - 42 a x^3 + 4 a^3 x^3 - 3 b
x^3 + 9 c x^3 + 4 a^2 x^4 + a x^5) y^2 + \\&& ~~ 14 a x (2 a + x)^3
(10 a + 9 x) y^3 - a (200 a^3 + 532 a^2 x + 486 a x^2 + 135 x^3)
y^4)\,.  \end{eqnarray*} Also in this case we cannot reach $\nu_7 =0$
identically by any choice of the parameters $a,b,c$. The existence of
a term which does not containg parameters allows us to conclude that
the other members of this series have the same property. Naturally,
all these formulas will gratly simplify in the case $a=0$. We can
expect, that the behaviour of the system (\ref{44}) in that case will
show some special features.

We take 
into account also the second projection of (\ref{44}), i.e., that on the $(x,
y)$-plane. The corresponding second order differential equation has the 
form (\ref{4}) and the coefficients $a_i(x,y),~i=1,...,4$ are given by:
\begin{eqnarray}
&&a_1=0,~~~a_2={1\over 3y},~~~a_3={ax+c-a-y\over 3y},\nonumber\\
&&a_4={ax^2+cx+b\over y^2}-{(a^2+1)x+ac-1\over y}+a,
\label{50}\end{eqnarray}
where $a,b,c$ are constants of the system (\ref{50}). The calculation of 
the semi invariant $\nu_5$ results in
\begin{eqnarray}
\nu_5&=&{-a+c+a x\over 9 y^{11}}(18 a^2 b - 36 a b c + 18 b c^2 + 
  (-36 a^2 b + 18 a^2 c + 36 a b c - 36 a c^2 +\nonumber\\&&
 ~~ 18 c^3) x + 
  (18 a^3 + 18 a^2 b - 72 a^2 c + 54 a c^2) x^2 + 
  (-36 a^3 + 54 a^2 c) x^3 + 18 a^3 x^4 + \nonumber\\&&
 ~~
  (4 a^4 + 45 a b + 27 a c - 16 a^3 c - 36 b c - 27 c^2 + 24 a^2 c^2 - 
     16 a c^3 + 4 c^4 +\nonumber\\&&
 ~~
 (54 a^2 - 16 a^4 - 36 a b - 36 a c + 
        48 a^3 c - 36 c^2 - 48 a^2 c^2 + 16 a c^3) x + \nonumber\\&& ~~
     (-9 a^2 + 24 a^4 - 72 a c - 48 a^3 c + 24 a^2 c^2) x^2 + 
     (-36 a^2 - 16 a^4 + 16 a^3 c) x^3 +\nonumber\\&&~~ 4 a^4 x^4) y + 
  (-12 a^2 c + 24 a c^2 - 12 c^3 + (-12 a^3 + 48 a^2 c - 36 a c^2) x +
\nonumber\\&& ~~
     (24 a^3 - 36 a^2 c) x^2 - 12 a^3 x^3) y^2 + \nonumber\\&& ~~
  (-a^2 - 8 a c + 8 c^2 + (-8 a^2 + 16 a c) x + 8 a^2 x^2) y^3)\,.
\label{45} \end{eqnarray} The projective curvatures $L_1$ and $L_2$
are given by \begin{eqnarray} L_1&=&{1\over
3y^4}(9(ax^2+cx+b)+2y(a-c)^2-4axy(a-c)+2a^2x^2y+\nonumber\\&&~~y^2(a-2c-
2ax)),\nonumber\\ L_2&=&{c-a+ax\over y^3}\,.\label{46}\end{eqnarray}
The semi invariant $\nu_5$ will be equal to zero only for $a=c=0$ with
arbitrary $b$. In this case the system (\ref{50}) has a very simple
form as well as the projective curvatures (\ref{46}): \[L_1={3b\over
y^4},~~~L_2=0. \] The calculation of the next possible series of
invariants gives us using formula (\ref{11a}) \[w_1=0.\] Consequently
we must turn over to the series with the first invariant beeing $i_2$
(\ref{15},\ref{16}). For this semi invariants the formulas hold
\begin{eqnarray*} i_2={15b\over y^5},~~i_4={135b^2\over
y^{10}},~~i_6={2430b^3\over y^{ 15}},\\ i_8={65610b^4\over
y^{20}},~~i_{10}={2361960b^5\over y^{25}},...  \end{eqnarray*} and the
corresponding absolute invariants (\ref{17}) are given by
\begin{equation} j_4={3\over 5},~~j_6={18\over 25},~~j_8={162\over
125},~~ j_{10}={1944\over 625},...\label{500}\end{equation} Because of
$L_{2}=0$ we cannot expect to find a first integral of the system in
the form of (\ref{18}). But we see, that in this case $V_2(x, y)$ can
be immersed into ${\Bbb P}^3(\Bbb R)$.

Let's mention that for $a=b=c=0$ both curvatures (\ref{46}) equal zero and 
that therefore our equation (\ref{50}) can be reduced to $y''=0$ and 
trivially solved.

Now we turn to the last case - the $(x,z)$-projection of the system 
(\ref{44}). The coefficients of the corresponding equation (\ref{4}) are
\begin{eqnarray*}
&&a_1=0,~~a_2={1\over 3y},~~a_3={b-3y-a\over 3ky},\\
&&a_4={a\over x}-{2(b+y)\over x^2}+{x+1\over y}+{cx-b-ax^2\over y^2}-{ab\over 
xy}.\end{eqnarray*}
The semi invariants $\nu_5$ has in this case the form
\begin{eqnarray*}
\nu_5={(ax-b)\over 9x^5y^{10}}(2b^4+ab^3x+9b^2x^2-6a^2b^2x^2+
x^3(-9ab+a^3b-9b^2+9bc)+\\  ~~~~~~~~~~~~
x^4(-9a^2+2a^4-18ab)+18a^2x^5-2y^2b^2+3abxy^2) \end{eqnarray*}

and the curvatures are
\begin{eqnarray*}
L_1&=&{-9bx^2-9cx^3-9ax^4+2b^2y-4abxy+2a^2x^2y-2by^2+3axy^2\over 3x^2y^4},\\
L_2&=&{b-ax\over xy^3}.
\end{eqnarray*}
We also investigated the invariants $\nu_5$, $\nu_7$, ... in the general 
case ($a,b,c\ne 0$). As an example we note here 
\begin{eqnarray*}
\nu_7&=& {1\over 27 x^7 y^{15}} (180 b^6 x^2 - 90 b^5 (a + 2 c) x^3 + 
    90 b^4 (9 - 7 a^2 + 2 a b + a c) x^4 + \\&&  ~~
    90 b^3 (-18 a + 7 a^3 - 9 b - a^2 b + 7 a^2 c) x^5 + 
    90 b^2 (a^4 - 7 a^3 b + 9 a c - 7 a^3 c + \\&& ~~ 9 b c - 9 c^2)
x^6 + 90 a b (9 a^2 - 2 a^4 + 18 a b + 7 a^3 b - 9 b^2 - a^3 c +
       18 b c + 9 c^2) x^7 + \\ && ~~
90 a^2  (-18 a b + a^3 b - 9 b^2 - 9 a c + 2 a^3 c - 45 b c) x^8 + \\ &&  ~~
    90 a^3 (9 a - 2 a^3 + 36 b + 18 c) x^9 - 1620 a^4 x^{10} + 
    10 (-b + a x)^3 (2 b^4 + \\&&  ~~
a b^3 x + 9 b^2 x^2 - 6 a^2 b^2 x^2 - 
       9 a b x^3 + a^3 b x^3 - 9 b^2 x^3 + 9 b c x^3 - 9 a^2 x^4 + \\&&  ~~
       2 a^4 x^4 - 18 a b x^4 + 18 a^2 x^5) y + 
    27 x (b - a x)^2 (a b^3 - 6 b^2 x + 5 a b x^2 - \\ && ~~ a^3 b x^2
- 3 b^2 x^2 + 11 b c x^2 - 20 a b x^3 + 2 a^2 x^4) y^2 + \\ && ~~
    14 b (-b + a x)^3 (-2 b + 3 a x) y^3 + 
    (-8 b^4 + 17 a b^3 x - 9 a^3 b x^3) y^4)\,.  \end{eqnarray*} We
can reach $\nu_5=0$ only for $a=b=0$. The semi invariant $\nu_7 =0$
only for $a=b=0$ in this case too as well as for all following
invariants of this series.  Then we have $L_1=3cx/y^4$, $L_2= 0$ and
$w_1=0$. The semi invariants of the series (\ref{15})-(\ref{16}) in
this case are \begin{eqnarray} &&i_2={15cx\over
y^5},~~i_4={135c^2x^2\over y^{10}},~~i_6={2430c^3x^3\over y^{
15}},\nonumber\\ &&i_8={65610c^4x^4\over
y^{20}},~~i_{10}={2361960c^5x^5\over y^{25}},...\, .
\label{54}\end{eqnarray} The corresponding absolute invariants are
exactly the same as in the former case (\ref{500}). In the considered
case $V_2(x,y)$ can be also immersed into ${\Bbb P}^3({\Bbb R})$.

\section {Conclusion}
We have investigated some projections of well known dynamical systems -- 
the Lorenz- and the R\"ossler-system. The proposed new approach to the 
investigation of those systems gives us a simple possibility to distinguish 
the regions of the parameters for which the systems show regular behaviour. 
Also, it turned out to be possible in some cases to find some
first integrals of the  
considered systems. These observation lets us hope that the method suggested 
can be used as a first step for the investigation of further new systems. 
Using the proposed in this paper method
 we can distinguish different areas in the space of parameters which 
allow a chaotic behaviour. This knowledge can be used as input for different, 
 regular or numerical for instance, methods. The present paper is
naturally a first step towards a more general approach which allows to
find an analytical criterium for the parameters of the system under
what the system has a certain dynamical state, a regular or a
stochastic one.

\section*{Acknowledgements}
The second author, V.S.D., thanks DAAD for financial support and the 
Mathematical Institute of Leipzig University for kind hospitality.
Both authors are deeply indebted the librarian, Mrs. I. Letzel for her  
effort to find the old and partly forgotten scientific papers.


\end{document}